\documentclass{article}
\usepackage{blindtext}
\usepackage[a4paper, total={6in, 8in}, lmargin=40pt, rmargin=40pt, tmargin=60pt, bmargin=60pt]{geometry}
\usepackage{graphicx}
\usepackage{tabularx}
\usepackage{booktabs} 
\usepackage{hyperref}
\hypersetup{hidelinks}
\usepackage[square,numbers]{natbib}
\usepackage{lineno}
\usepackage{amsmath, xparse}

\begin{document}

\title{Integrating Conventional Headway Control with Reinforcement Learning to Avoid Bus Bunching}
\author{Xiheng Wang}
\date{}

\maketitle
\renewcommand{\abstractname}{\large\bf Abstract}
% Abstract (Do not insert blank lines, i.e. \\) 
\begin{abstract}
    Bus bunching is a natural-occurring phenomenon that undermines the efficiency and stability of the public transportation system. The mainstream solutions control the bus to intentionally stay longer at certain stations. Existing control methods include conventional methods that provide a formula to calculate the control time and reinforcement learning (RL) methods that determine the control policy through repeated interactions with the system. In this paper, we propose an integrated proximal policy optimization model with dual-headway (IPPO-DH). IPPO-DH integrates the conventional headway control with reinforcement learning, so that it acquires the advantages of both algorithms --- it is more efficient in normal environments and more stable in harsh ones. To demonstrate such an advantage, we design a bus simulation environment and compare IPPO-DH with RL and several conventional methods. The results show that the proposed model maintains the application value of the conventional method by avoiding the instability of the RL method in certain environments, and improves the efficiency compared with the conventional control, shedding new light on real-world bus transit system optimization.
\end{abstract}

% Keywords
\vspace{3em}

\pagebreak

\renewcommand{\contentsname}{\large\bf Content}
\tableofcontents 

\renewcommand{\figurename}{\large\bf Figure}
\renewcommand{\tablename}{\large\bf Table}
\renewcommand{\refname}{\large\bf Bibliography}
%%%%%%%%%%%%%%%%%%%%%%%%%%%%%%%%%%%%%%%%%%
\section{Introduction}
\begin{figure}[!htbp]

\includegraphics[width=1.0\linewidth]{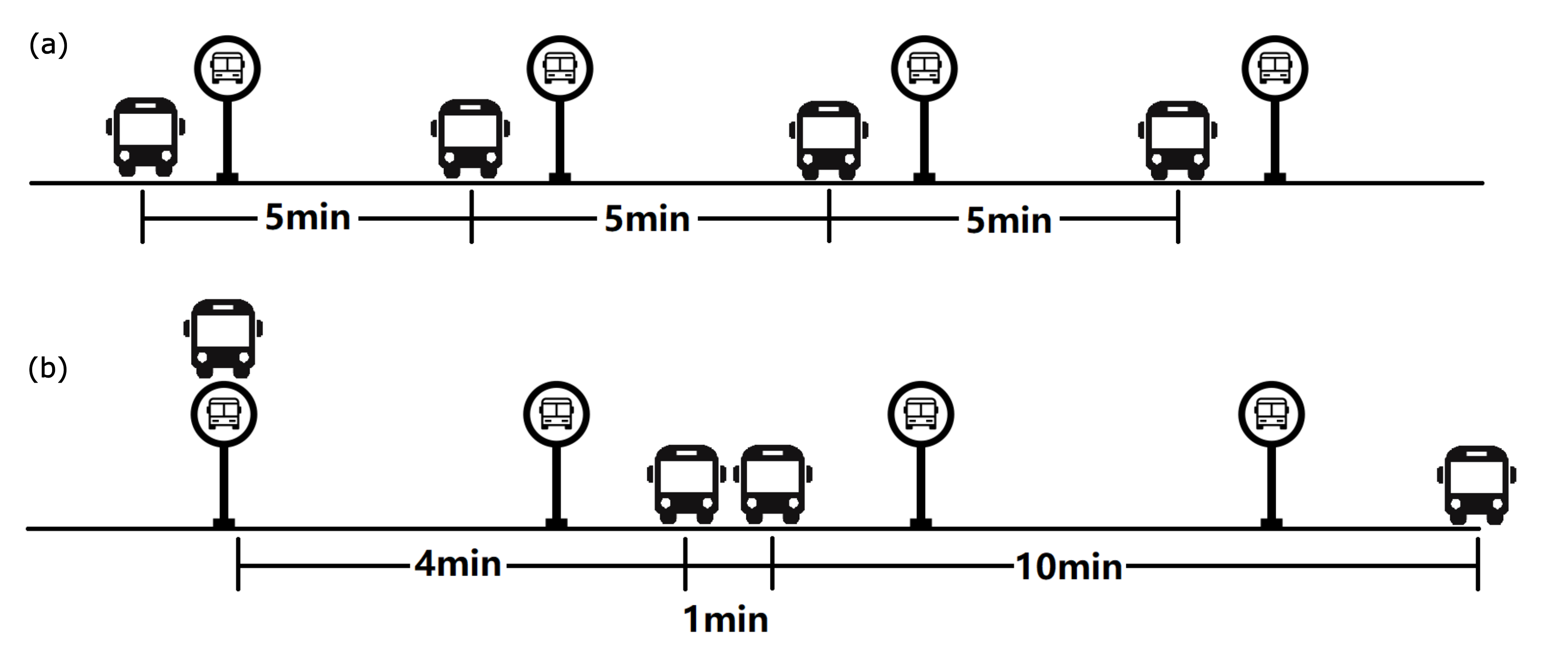}
\caption{Illustration of bus bunching. 
(\textbf{a}) Buses operate with even headways of roughly 5 minutes. 
(\textbf{b}) Two buses bunch together, causing uneven headways in the system.
\label{busbunching}}
\end{figure}

Public bus transportation is a core component of a city’s comprehensive transportation system. It also plays a pivotal role in achieving carbon neutrality, supporting the sustainable development of future cities. With the increasing number of vehicles on the road in recent decades, the bus system has become increasingly fragile, hindering the adoption of buses in cities. During my daily commutes between school and home, I sometimes encounter a rather irritating phenomenon - waiting at the bus stop for 10 minutes, three times the usual bus interval, and then seeing two or three buses come into the station together. Bus bunching, as the phenomenon is called, is where two or more buses arrive at the same stop at the same time. It is one of the problems that heavily diminishes the transit efficiency of the system. The causation of bus bunching includes not only the driving behavior of bus drivers but also the traffic lights and congestion in the route. Bus bunching occurs naturally in an uncontrolled bus system, even in small cities with light traffic. Once the natural disturbance of the system causes the late arrival of buses, there will be more passengers waiting at the bus stop. Consequently, we can expect that the bus needs more time to load the passengers, which further delayed the running of the bus system. This creates a positive feedback loop that will eventually cause several buses in an uncontrolled system to bunch and travel together after a period (Figure \ref{busbunching}).

\noindent Generally, bus bunching has two major negative impacts on the transit system.

\begin{itemize}
\item   Passengers will be waiting for a longer period of time and spend more additional time to reach their destinations owing to the longer holding time of the bus at stops, which would accelerate the transfer of passengers from public transportation to driving or taxi.
\item	Bus bunching disrupts the normal distribution of passengers waiting at the bus stops, which further lowers the transit efficiency of the bus systems and indirectly leads to more energy consumption and exhaust pollution per capita.
\end{itemize}

Overall, bus bunching undermines the efficiency of public transportation systems and lowers the citizen’s satisfaction, posting a significant challenge for large-scale cities with transit demand on the rise. As the information age arrives, algorithm-based intelligent transportation systems (ITS) are gradually becoming new tools to solve those problems in big cities. The mainstream method of control requires that the target buses hold at selected bus stops even after all of the passengers have been onboard, in order to recover the initial distance between two buses. This solution explores the trade-off between temporarily stopping the bus and the occurrence of bus bunching. In this context, most solutions need to answer two questions: 
(\textbf{i}) Which bus needs to be controlled? 
(\textbf{ii}) How much additional time should each bus hold at each stop?
The optimization target always is the stability of the running of buses, represented by the variance of distances between neighboring buses. In light of this, several researchers proposed different algorithms to eliminate bus bunching with a suitable control period at certain stops. 

Conventional methods proposed by transportation experts include schedule-based algorithms~\cite{boyle2009, abkowitz1990, van2010}, single-headway control algorithms~\cite{dgz2009,xuan2011, bartholdi2011} and dual-headway control algorithms~\cite{dgz2011}, all of which calculate the holding time with static formulas. Those conventional methods are usually stable in a wide range of environments with different passenger arrival densities, but there is still space for higher efficiency. With the advance in machine learning methods, bus bunching can be dealt with reinforcement learning (RL) algorithms. On this track, Chen et al.~\cite{chen2016} and Alesiani et al.~\cite{alesiani2018} adopt a MARL framework; Wang et al.~\cite{wang2020} proposed a reward function to promote headway equalization. These model-free networks push the efficiency of the system to its upper limit and successfully eliminate the bus bunching problem. Nevertheless, a big concern with RL methods is their instability. An unexpected large disturbance might drive the system outside the familiar normal training range and break the system’s balance, leading to disrupted transit systems. This inability also keeps RL-based algorithms outside real-world applications, since stability is a major safety concern for systems operating in real life.

In this paper, we make three further contributions relative to the state-of-the-art in the field:
\begin{enumerate}
\item	We analyze the existing methods under extensive simulations and discuss the situations they each perform well in.
\item	Building on the limitations of the existing algorithms, we propose an integrated method IPPO-DH that gathers the advantage of both the existing RL and non-RL control;

\item	We compare the IPPO-DH with the existing methods under different simulation environments to confirm its value at increasing the efficiency without sacrificing the ability to stabilize the system under all settings.
\end{enumerate}

\noindent \space

The rest of this paper is structured as follows:
In section~\ref{s_EW}, we review the existing method of control. Section~\ref{s_M} introduces the new methodology proposed in this paper. Section~\ref{s_E&E} presents the experiments and result analysis on the new method and compares it with existing ones. Section~\ref{s_D} summarizes this work.

%%%%%%%%%%%%%%%%%%%%%%%%%%%%%%%%%%%%%%%%%%
\section{Related Work} \label{s_EW}

Several holding control algorithms exist to deal with the bus bunching problem. For traditional, non-RL, methods, Zhao et al.~\cite{zhao2006} provided an optimal slack time for pre-determined schedule control that minimizes the passenger waiting time at the station. Xuan et al.~\cite{xuan2011} proposed a virtual schedule method that further explores the static control domain and increases the efficiency of the schedule-based method. Other than static control methods, Daganzo~\cite{dgz2009,dgz2011} introduced a dynamic control method that determines control time with forward headway, and is followed by another method that controls the bus by both the forward and the backward headway.

With the advance in machine learning theory and technology, reinforcement learning can play a key role in solving this problem. Several multi-agent reinforcement learning models were established in recent years. Wang et al.~\cite{wang2020} proposed a multi-agent deep reinforcement learning method that outperforms the traditional methods. They proposed a multi-agent PPO with a joint action tracker framework to address the problem. Foerster et al.~\cite{foerster2018} proposed a multi-agent actor-critic algorithm with a counterfactual baseline for better credit assignment with a discrete action space. Sun et al.~\cite{wang2021} proposed a credit assignment framework for asynchronous control that better characterized the marginal contribution of other agents during the policy training process.

However, the conventional control methods are usually skilled in the stability in solving bus bunching problems, while the RL-based methods are able to achieve better performance with weak stability. To this end, in this paper, we couple a modified multi-agent PPO reinforcement learning with the dual-headway control method into one framework, such that the framework could balance the efficiency and stability of bus control in complex environments.

\subsection{Headway-based control}
Dual-headway approach was proposed by Daganzo in 2011~\cite{dgz2011}, marking the best conventional (non-RL) solutions to the problem. This dynamic control method sets the control time according to the forward headway and backward headway of the bus. The real-time speed of each bus is given by

\begin{linenomath}
\begin{equation}
{\nu}_{n}=\overline{\nu}+{C_n}(\mathbf{h},{S}),\quad  n \in \{1, 2, \dots, N \} ,
\label{eqn_DualHeadwaySpd}
\end{equation}
where $\mathbf{h}$ is the headway vector, $\mathbf{h}{\in}\mathbf{R}^{N}$. $S$ is the standard headway of the system in equilibrium. $\overline{\nu}$ is the ideal maximum speed of the bus with no external disturbance. Both $S \in \mathbf{R}$ and $\overline{\nu} \in \mathbf{R}$ are constants.
${C_n}(\mathbf{h},S)$ is further expanded into
\begin{equation}
{C_n}(\mathbf{h},{S})=\frac{[\lambda{b}\overline{\nu}(h^-_n-S)+{\alpha}(h^-_n-h^+_n) -{\delta}]}{(1-{\lambda}b h^-_n)},\quad  n \in \{1, 2, \dots, N \}, 
\end{equation}
where $\lambda {\in} \mathbf{R}$ captures the arrival rate of passengers (pax) with unit $\mathrm{pax}/(\mathrm{km}\cdot \mathrm{h})$, $b {\in} \mathbf{R}$ is the time bus meet a passenger, unit $\mathrm{h}/\mathrm{pax}$. $h^-_n$ denotes the forward headway and $h^+_n$ denotes the backward headway of the $n^{th}$ bus, $h^-_n, h^+_n {\in} \mathbf{R}$.  $\alpha, \delta \in \mathbf{R}$ are free constants.
\end{linenomath}

This method is bringing more challenges to implement on an operational level due to the need for both headway information, but, in turn, yields better efficiency compared to static control. 

%%Turn S into H, denoting headway,
In Equation~\ref{eqn_DualHeadwaySpd}, 
in order to maximize the velocity ${\nu}_{n}$, we need to max out ${C_n}(\mathbf{s},{S})$, which is a minimization task for constant $\delta$. The calculation given in the paper~\cite{dgz2011} derived the minimum $\delta$ to be 
$
\delta^{\ast} \approx 5.0 r (\lambda b \overline{\nu})^{\frac{1}{2}},
$
when
$
 \alpha^{\ast} \approx 0.63\lambda b \overline{\nu}.
$

Following Berrebi et al.~\cite{berrebi2018}, we transform the continuous holding method into equivalent control time at stations as:
\begin{equation}
a_n = (\alpha+\beta)(S-h^-_n) - \alpha(S-h^+_n), \quad n \in \{1, 2, \dots, N \}, 
\label{eqn_DualHeadwayControlTime}
\end{equation}

where $\alpha$ is a constant between 0 and 1, and can be seen as the weighting term between forward and backward headway control.  $\beta$ denotes the delay of a bus resulting from a unit headway increase, determined by the system.

\subsection{Reinforcement Learning Method}
The Multi-agent PPO Reinforcement Learning actor-critic framework is based on the traditional actor-critic framework proposed by Sutton et al.~\cite{sutton1998}.
Specifically, the decision-making of each agent in the environment is based on local information. In particular, the agent uses the forward and backward headway to determine the holding time through its "actor" component (usually a neural network) and receives a reward given by the environment. The "critic" component is responsible for estimating the state value given global information (i.e. the system headways and the other agents' actions). The state value estimation is then used by the actor to update its policy. Wang et al.~\cite{wang2020} point out that limiting actors to local information can be beneficial in avoiding extra approximation during real-world application.

\subsection{Weakness of Existing Work}
Using the information collected from the simulation described in section~\ref{s_E&E}, we analyze and compare the comparative advantage and limitations of each method.
The conventional method is more reliable in all different environments compared to the RL method. The lower failure rate makes it preferable for real-world solutions. On the other hand, by adopting the RL method of control, even though the overall system efficiency would be largely improved, the instability of the system under control would increase, leading to more risk in real-world applications. To create a solution that contains the stability of the conventional method and the high performance of the RL method, we integrated the two methods together.

Via comprehensive experiments in our simulation environment, we confirm that a normal RL method would not work in extreme situations of high passenger arrival. To tackle this issue, we couple the dual-headway formula control method with RL, to enhance the reliability of a simple RL algorithm in the face of complex situations. Specifically, we train an RL network to update the decision of the dual-headway method with a delta amount to promote the performance of the algorithm in cases of low or moderate passenger arrival rate. In this way, both advantages, the stability of the conventional method and the efficiency of the RL model, can be integrated together.

%%%%%%%%%%%%%%%%%%%%%%%%%%%%%%%%%%%%%%%%%%
\section{Methodology}\label{s_M}
Following the idea of combining two methods, we propose the Integrated PPO Model with Dual-headway (IPPO-DH).

%\noindent \textbf{Overall Framework}

\begin{figure}[!htbp]
\includegraphics[width=1.0\linewidth]{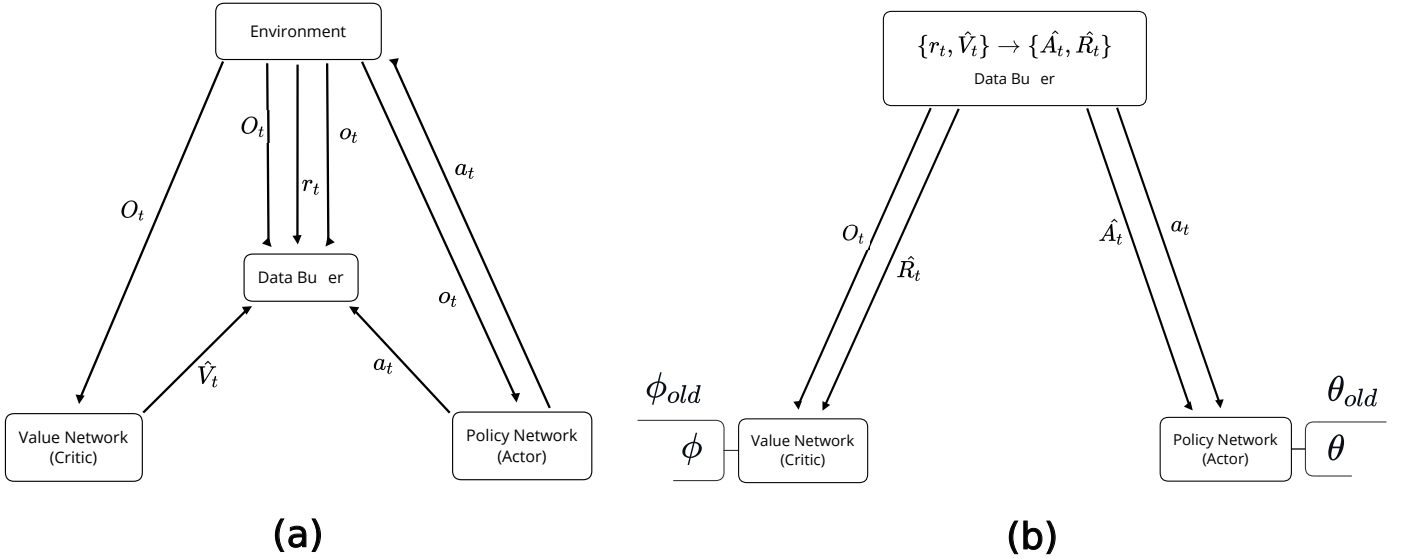}
\caption{Framework of the IPPO-DH model.  (\textbf{a}) Run the simulation to collect trajectories. (\textbf{b}) Process data and train model. \label{overall_framework}}
\end{figure}

The method is a combination of reinforcement learning (RL) with conventional dual-headway control. The system entities include the environment and the agent, which consists of a value network (critic) and a policy network (actor) (Figure~\ref{overall_framework}). We assume that all buses in the system are homogeneous, so they shared one set of parameters. 
%The training process can be divided into two stages. We will now explain the first stage and the definition of basic components (state, action, reward), before considering the second stage and the training process (loss functions). 
Figure~\ref{overall_framework} describes the flow of information in the system. $O_t$ denotes the global information, $o_t$ denotes local information, $a_t$ is the action performed by the agent, and $r_t$ is the reward. In the RL module, the environment passes on global information ($O_t$) to the critic and local information ($o_t$) to the actor. The actor performs an action ($a_t$), and the environment reacts by calculating a reward ($r_t$). 

Apart from the normal actor-critic method, one critical issue for the bus environment is the asynchronous control timing among agents. Therefore, instead of adopting online training, we use the off-policy version of actor-critic by setting up a replay data buffer to store the state, action, reward, and time information. 
$\phi,\ \theta$ denotes the parameters of the value and policy networks. $\phi_{old},\ \theta_{old}$ are copies of the network parameters ($\phi$, $\theta$) and are updated after each epoch.

The basic components of our RL framework are designed as follows:
\begin{itemize}

\item \textbf{State:}

\begin{itemize}
\item \textbf{Policy (Actor) State:} $S_{actor}=[h^-_t,h^+_t]$, a $\mathbf{R}^2$ vector contains the local observation of the agent: $h^-_t$ denotes the forward headway and $h^+_t$ denotes the backward headway.

\item \textbf{Value (Critic) State:} $S_{critic} = (H_t,\pmb{\nu_t})$. $H_t =  \begin{bmatrix}h_1 & h_2 &  \cdots & h_N \end{bmatrix} $ is the headway vector between all buses, $H_t\mathrm{\in}\mathbf{R}^{N}$. 
The $n^{th}$ component always represents the headway between the ${(n-1)}^{th}$ and the $n^{th}$ buses ahead of the agent. $ \pmb{\nu_t} =  \begin{bmatrix} {\nu}_1 & {\nu}_2 &  \cdots & {\nu}_N \end{bmatrix} $
is the responsibility vector, $\pmb{\nu_t}\mathrm{\in}\mathbf{R}^{N}$. 
The $i^{th}$ component $\nu_i$ represents the relative holding percentage of the $i^{th}$ bus ahead of the agent during the holding action of the agent itself. The intuition is, that the value of a state is not only dependent on the headway of the system but also on the action others are taking. When training with the responsibility vector added, the value estimation takes the effect of simultaneous action within the system into account, thus promoting bus-to-bus cooperation in making more accurate action decisions during the training process.
\end{itemize}

\item \textbf{Action: }

% \noindent \textbf{Integrated Actor Agent}
To combine the advantage of both the conventional and RL control method, we redesign the integrated action to be 
\begin{linenomath}
\begin{equation}
a_t =\mathrm{clip}( C + \Delta a \cdot H_{max}, 0, H_{max}).
\label{eqn_IntegratedAction}
\end{equation}
\end{linenomath}

The output of the policy network is a strength parameter $\Delta a \in [-1,1]$. $C$ is the conventional holding time provided by the dual-headway method given by  Equation~\ref{eqn_DualHeadwayControlTime}, $C \in \mathbf{R}$. $H_{max} \in \mathbf{R}$ is a constant which is manually set to denote the maximum holding time for a bus at a certain bus stop.  
The intuition is to ask the RL network to revise the output of the conventional method in a more stable environment (less passenger arrival) to increase system efficiency while keeping the stability of the environment under severe circumstances with large passenger arrival.

\item \textbf{Reward: }
The agent’s performance can be assessed in two aspects: Headway equalization and holding time. Hence, we follow the two-term reward function proposed by Wang et al.~\cite{wang2020}. This reward function (Equation~\ref{eqn_Reward}) is designed to avoid the task of crediting rewards among different agents since it depends only on the local state the single agent is in. 

%%Note: put w1 and w2 in this formula instead of a single w!
\begin{linenomath}
\begin{equation}
r_t = \omega_1 \cdot \exp(-|h^{-}_t - h^{+}_t|) + \omega_2 \cdot \exp(-a_t).
\label{eqn_Reward}
\end{equation}
\end{linenomath}

$a_t \in \mathbf{R}$ denotes the action holding time at time $t$, given by Equation~\ref{eqn_IntegratedAction}.
The first term represents the degree of headway equalization, while the second term encourages less holding time. It would be the bus’s task to maximize the reward and balance a trade-off between intervening the environment less and keeping it stable. $\omega_1$ and $\omega_2$ are strength hyperparameters that are set manually, $\omega_1 \in \mathbf{R}, \omega_2 \in \mathbf{R}$.
\end{itemize}
\smallskip

%%%%%%%%%%%%%
%% Stage 2 %%
%%%%%%%%%%%%%
After each epoch of simulation, we then perform the off-policy update on both value and policy networks.
We first transform the reward and value estimation sequences $\left\{ r_t,V_t \right\}$ into Return sequence $\left\{ \hat{R_t} \right\}$ and advantage sequence $\left\{ \hat{A_t} \right\}$ according to generalized advantage estimation (GAE)~\cite{gae}, and then use the $\left\{ \hat{R}_t \right\}$ for value network training and $\left\{ \hat{A}_t \right\}$ for policy network training

\noindent \textbf{Lowering Training Bias by using PPO on policy training:}

Since our data is collected with the old policy and value estimation (off-policy), we keep introducing more bias as we train the network off-policy using batch data (i.e. $\theta\leftarrow \theta+\alpha\cdot {\nabla}_\theta  J(\theta)$). This is because when the difference between the old and new policy ($\pi_\theta$) increases, so does the inaccuracy of value estimation. The bias usually causes the training process to be less stable. To solve this, we use proximal policy optimization (PPO) proposed by Schulman et al.~\cite{schulman2017}. PPO sets a restriction for the induced bias by restricting the policy change in a single training episode. The objective function of PPO is

\begin{linenomath}
\begin{equation}
J_{PPO}\left(\theta\right) = E_{a\sim \pi_{\theta_{old}}} 
\left[ 
    \min \left( 
        \delta\left(\theta\right) \hat{A}, \mathrm{clip} \left(\delta\left(\theta\right),1-\epsilon,1+\epsilon\right) \hat{A}
    \right) 
\right].
\label{eqn_PPOLoss}
\end{equation}
More specifically, we replace the usual gradient descend objective functions with a clip variable in Equation~\ref{eqn_PPOLoss}. $\hat{A} \in \mathbf{R}$ denotes the advantage function of action in a state, and $\theta$ represents the parameters of the policy network. For a good action ($\hat{A}>0$), the policy ($\pi_\theta$) is updated only when the ratio of probability between the old policy network and post-training policy($\delta(\theta)=\frac{\pi_\theta(a_t|s_t)}{\pi_{\theta}^{old}(a_t|s_t)} \in \mathbf{R}$) is smaller than $1+\epsilon$. For bad actions ($\hat{A}<0$), the policy is updated when the ratio($\delta$) is larger than $1-\epsilon$. The hyperparameter $\epsilon$ controls the degree of changes the policy is able to have during one training epoch, which regulates the speed and enhances the robustness of the training process.
\end{linenomath}

For our policy network, its function is to obtain cumulative reward as much as possible under the advantage estimation ($\hat{A}$) given by the value network. Thus, the loss function to minimize is designed to be:
  
\begin{linenomath}
\begin{equation}
L_{policy}\left(\theta\right) = - \sum_{t=0}^{T}
\min \left( 
    \delta_t\left(a_t;\theta\right) \cdot \hat{A}_t^{\mathrm{GAE}(\gamma, \tau)}, \mathrm{clip} \left(\delta_t\left(a_t;\theta\right),1-\epsilon,1+\epsilon\right) \cdot \hat{A}_t^{\mathrm{GAE}(\gamma, \tau)}
\right),
\label{eqn_ActorLoss}
\end{equation}
\end{linenomath}
\noindent where the advantage $\hat{A}^{\mathrm{GAE}(\gamma, \tau)}$ is given by generalized advantage estimate (GAE)~\cite{gae},

\begin{linenomath}
\begin{equation}
\hat{A}^{\mathrm{GAE}(\gamma, \tau)} _t= \sum_{l=0}^{\infty}
\left( \gamma \tau \right)^{l} 
\cdot 
\left( r_t + \gamma \hat{V}_{t+l+1} - \hat{V}_{t+l} \right).
\label{eqn_GAEAdvantage}
\end{equation}
\end{linenomath}

\noindent
$\delta(a_t;\theta)$ in Equation~\ref{eqn_ActorLoss}
denotes the ratio of change between the old and new policy, 
$\delta(a_t;\theta)=\frac{\pi_\theta(a_t|s_t)}{\pi_{\theta}^{old}(a_t|s_t)}$. $\theta$ denotes the parameters of the value network, and $\epsilon$ denotes the clip hyperparameter described in PPO.

%%[critic loss] 
As the state value is constant regardless of the policy, we can directly train the value network by constructing a data set with the simulation trajectory:
$
\left\{ \left( 
s_t,\ \hat{R}^{\mathrm{GAE}(\gamma, \tau)}_t
\right) \right\},
$
and train the network with supervised regression:
\begin{linenomath}
\begin{equation}
L_{value}\left(\phi \right) = \frac{1}{2} \sum_{t=0}^{T}
\left|\left|
\hat{V}^{\phi, \pi}_t - \hat{R}^{\mathrm{GAE}(\gamma, \tau)}_t
\right|\right|
,
\label{eqn_CriticLoss}
\end{equation}
\end{linenomath}
where $\phi$ denotes the value network parameters, and
\begin{linenomath}
\begin{equation}
\hat{R}^{\mathrm{GAE}(\gamma, \tau)}_t = 
\hat{A}^{\mathrm{GAE}(\gamma, \tau)} _t + \hat{V}^{\phi, \pi}_t
.
\label{eqn_GAEReturn}
\end{equation}
\end{linenomath}

%%%%%%%%%%%%%%%%%%%%%%%%%%%%%%%%%%%%%%%%%%
\section{Experiment and Evaluation} \label{s_E&E}
\subsection{Environment Setting}
We set up the simulation environment as follows:
A total of $N=16$ buses operate on a two-way route back and forth. The buses operate at a constant speed $v_i \in \mathbf{R}$ on each segment, $v_i$ follows a normal distribution $v_i \sim {N}(27,4.5^2)$ for each segment (with a mean of $27$ km/h and a standard deviation of $4.5$ km/h) to simulate the uncertainty encountered between bus stations.
The route is $L= 9.21$ km in length, consisting of $M=17$ bus stops and 9 intersections distribute unevenly. This design corresponds geographically to a real-world bus running route in a big city.

\begin{figure}[!htbp]
\includegraphics[width=1.0\linewidth]{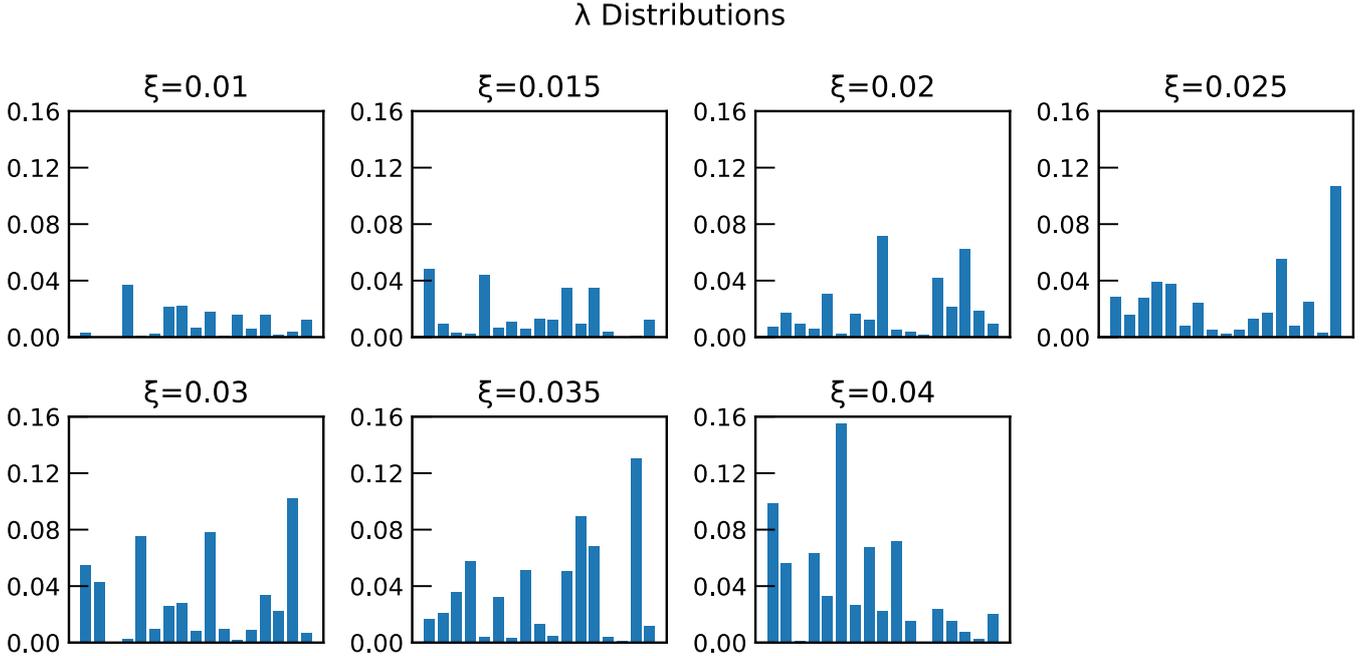}
\caption{The distribution of Poisson parameter $\lambda$ under different environments.
\label{Lambda}}
\end{figure}

Passengers are spawned randomly according to a Poisson distribution at each stop. The Poisson parameter $\lambda$ denotes the average arrival rate of the passengers for each stop with unit pax/minute.
We design seven simulation environments with different passenger arrival distributions by controlling the mean and standard deviation of the Poisson distributions of all stations. The resulting distributions are shown in Figure~\ref{Lambda}. In particular, we define $\Lambda = [\lambda_1,\lambda_2,\cdots,\lambda_M]$, and randomly set $\Lambda$ with the restrain of $ mean(\Lambda)=std(\Lambda)$. We let $\xi=mean(\Lambda)=std(\Lambda)$ representing the passenger density. A larger $\xi$ denotes a less stable environment, making it relatively harder for algorithms to avoid bus bunching. The time of boarding and alighting a bus is set to $t_b = 3\ \mathrm{s}/\mathrm{pax}$ and $t_a = 1.8\ \mathrm{s}/\mathrm{pax}$.
Moreover, Table~\ref{tab_hyperparam} shows the hyperparameters we used in the simulation.

%  figure $\xi=0.1 $ figure $\xi=0.2 $ figure $\xi=0.3 $ 
% \begin{figure}[H]
% \includegraphics[width=10.5 cm]{figures/lambda.eps}
% \caption{The distribution of Poisson parameter $\lambda$ under different environments. Under each environment, stations are sorted from the smallest $\lambda$ to the largest $\lambda$.
% \label{Lambda_notinuse}}
% \end{figure}

\begin{table}[h] 
\caption{The final hyperparameters chosen after tuning. \label{tab_hyperparam}}
\begin{tabularx}{\textwidth}{ | >{\centering\arraybackslash}X | >{\centering\arraybackslash}X |}

\hline
\textbf{Parameter}	& \textbf{Value}	\\
\midrule
\hline
$H_{max}$\textsuperscript{1}		& 180 	\\
\hline
$\omega_1$& 0.6 	\\
\hline
$\omega_2$& 1.0 	\\
\hline
$\gamma$& 0.8 	\\
\hline
$\tau$& 0.95 	\\
\hline
$\epsilon$& 0.5\\
\hline
\bottomrule
\end{tabularx}
\noindent{\footnotesize{\textsuperscript{1} $H_{max}$ has unit second.}}
\end{table}

\subsection{Key Metrics}
To compare the performance of the integrated method (IPPO-DH), we use the Dual-headway method and the pure RL method as baseline models. The primary goal of optimizing the transportation system is to increase passenger satisfaction. Thus, we choose the passenger waiting time and passenger travel time to be the key metrics that determine the model's performance. Passenger waiting time and passenger travel time are the only passenger-related variable that affects passenger experience directly. 
To further consider the feasibility of each method in practice, we also calculate the variance of the metrics throughout the simulation. Overall, the goal is to lower both the mean (efficiency) and variance (stability) of passenger waiting time and passenger travel time.

\subsection{Result and Analysis}
We test each of the proposed and baseline models during a 24-hour simulation and gather the consecutive locations of each bus at all time steps as well as the key metrics (e.g., passenger waiting time, passenger travel time) of every passenger after they finish their trips. In addition, we also gathered the suggested additional holding time for all control actions. We then calculated the mean and variance of the three variables and repeated the process in seven different environments described in Figure~\ref{Lambda}.

We will now analyze the two baseline methods (dual-headway control and RL control) in two representative environments, a stable one with $\xi=0.02$ and a harsh one with $\xi=0.04$. Then, we will compare our IPPO-DH method against both baseline methods in all seven environments regarding the key metrics.

%%%%%%%%%%%%%%%%%%%%%%%%%%%%
\subsubsection{Headway-based Control}
\begin{figure}[!htbp]
\includegraphics[width=0.48\linewidth]{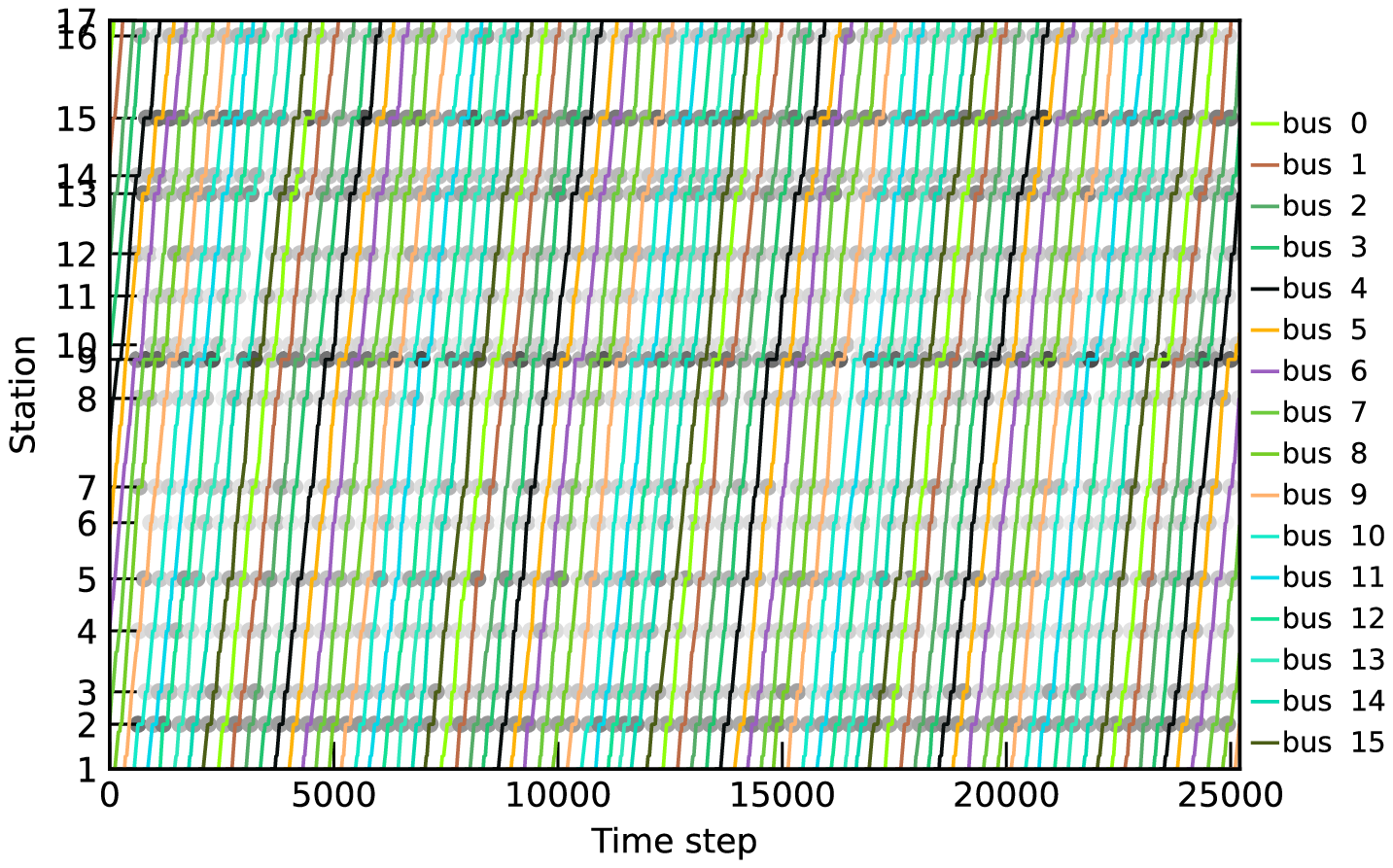}
\includegraphics[width=0.48\linewidth]{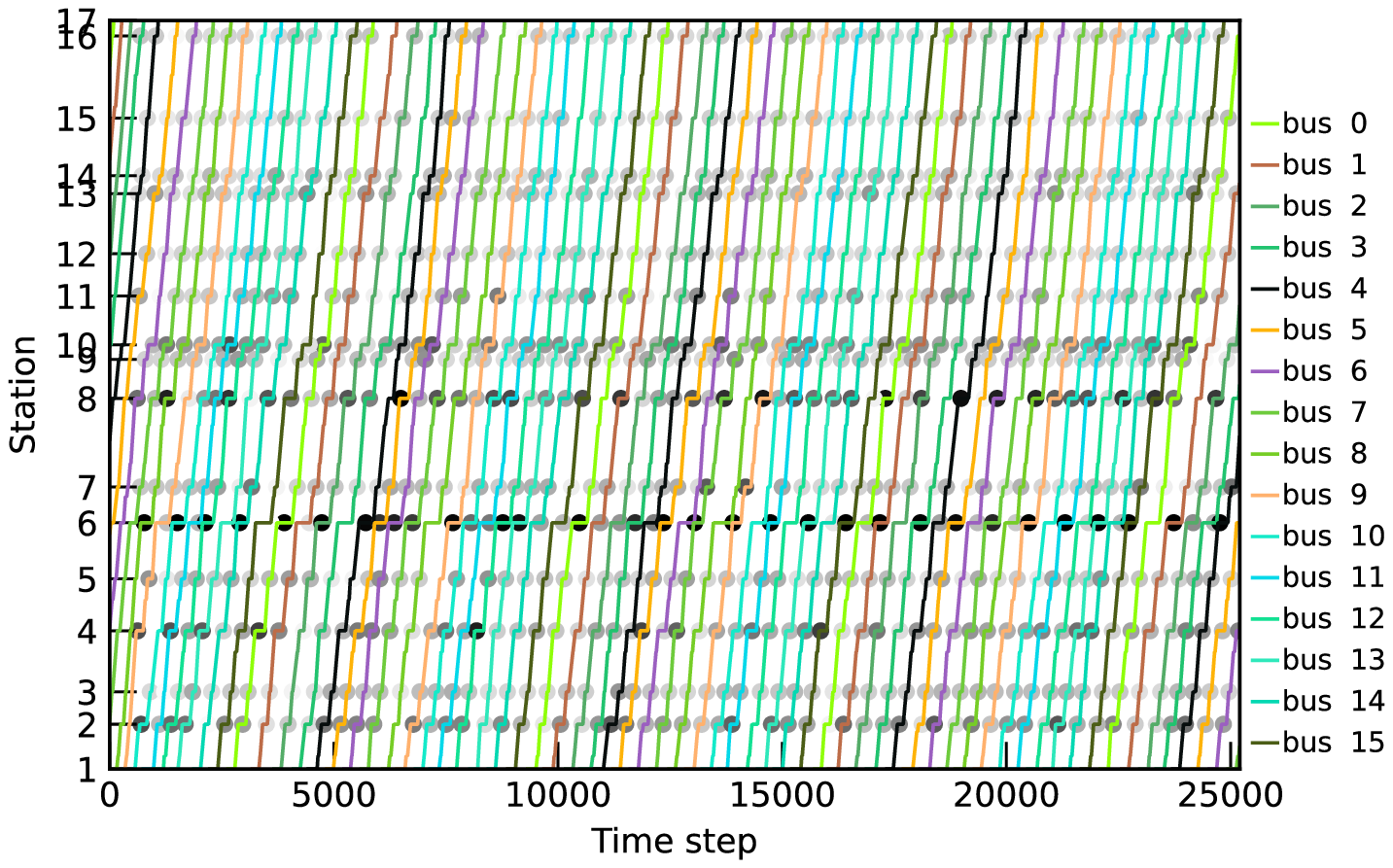}
\caption{The space-time diagram of the buses under Dual-headway control. (\textbf{a}) Simulation result in a stable environment ($\xi = 0.02$). (\textbf{b}) Simulation result in a harsh environment ($\xi = 0.04$). \label{ST_DH}}
\end{figure}

Running the dual-headway method in the simulation, we find that its performance is very stable among all simulations (Figure~\ref{ST_DH}). This is due to the constant restoring force acting on the system regardless of the external environment and current situation. Overall, this algorithm can operate in a larger range of environments. In a harsh environment with a high and uneven passenger arrival rate across stations, bus bunching can still be avoided with a slightly higher control time determined by Equation~\ref{eqn_IntegratedAction}. Both these attributes are very valuable in terms of real-world application.
 
\subsubsection{Reinforcement Learning Method}
\begin{figure}[h]
\includegraphics[width=0.48\linewidth]{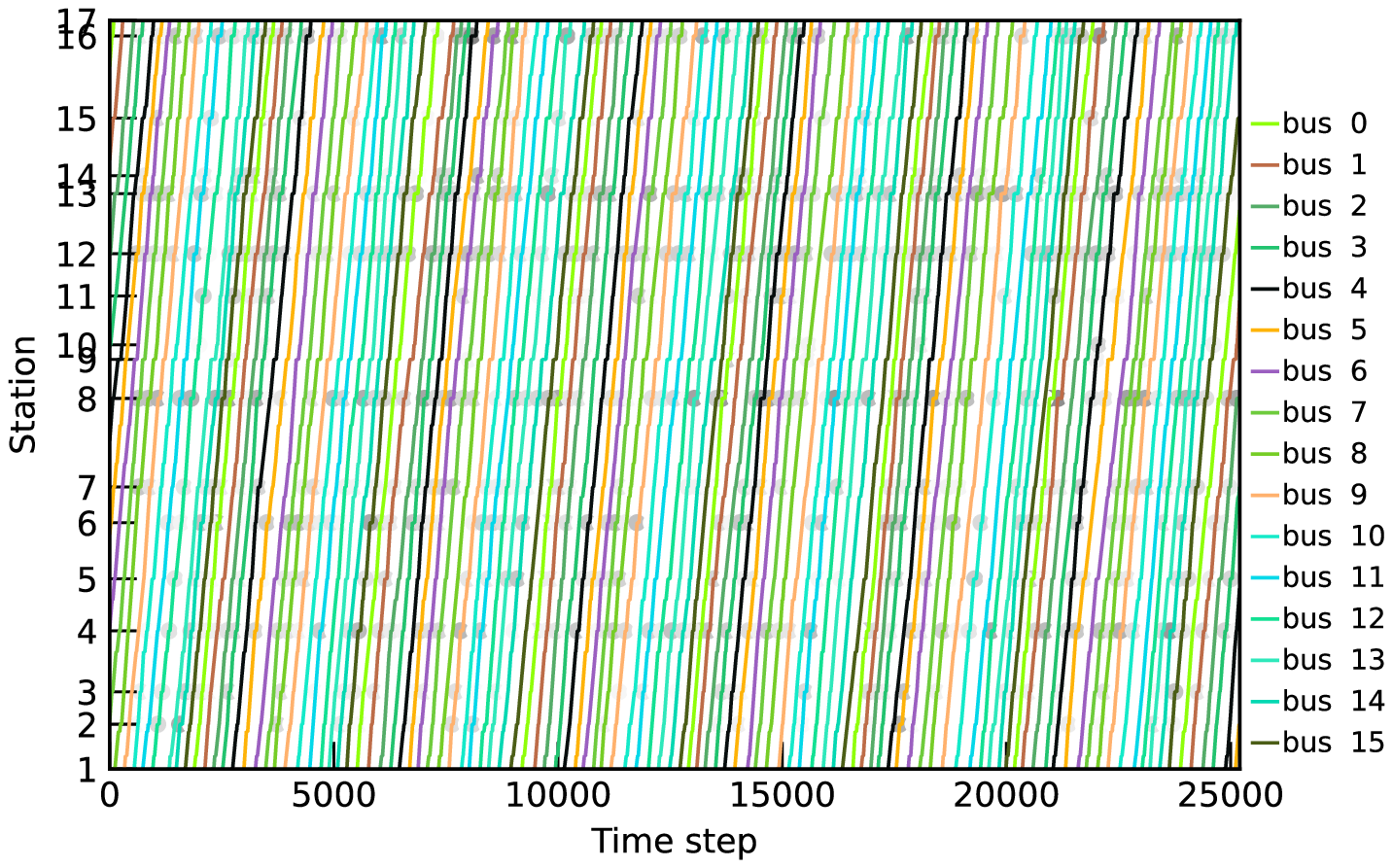}
\includegraphics[width=0.48\linewidth]{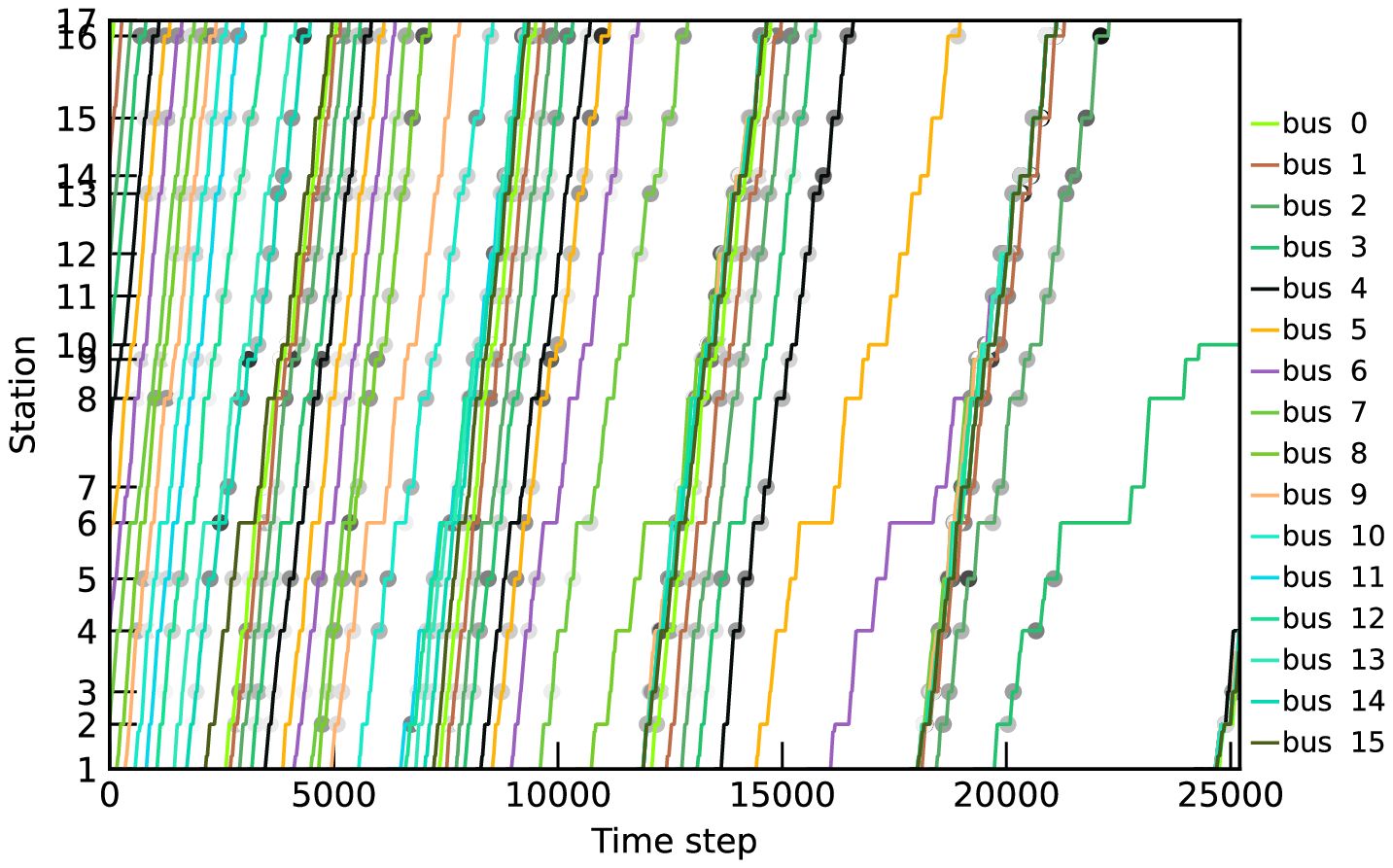}
\caption{The space-time diagram of the buses under RL control. (\textbf{a}) Simulation result in a stable environment ($\xi = 0.02$). (\textbf{b}) Simulation result in a harsh environment ($\xi = 0.04$). \label{ST_RL}}
\end{figure}

For the pure RL method, we trained a basic multi-agent PPO actor-critic framework within an environment with a lower-than-normal passenger arrival rate. When we set the arrival rate of passengers to normal ($\xi = 0.02$, Figure~\ref{ST_RL} \textbf{a}), the model showcases an outstanding performance with respect to the stated metrics. However, when the environment was made more challenging ($\xi = 0.04$, Figure~\ref{ST_RL} \textbf{b}), we notice that the model’s performance become more unstable. It became unstable in the middle of the simulation process, and lead to severe bus bunching. We suspect it to be an unexpected surge of passenger arrival at a certain station that breaks the balance. Despite the lack of transferability of an RL model between environments, its performance is still excellent when the environment is in a normal range. 

%%%%%%%%%%%%%%%%%%%%%%%%%%%%

\subsubsection{Integrated PPO Model with Dual-headway baseline (IPPO-DH)}

\begin{figure}[!]
\includegraphics[width=0.8\linewidth]{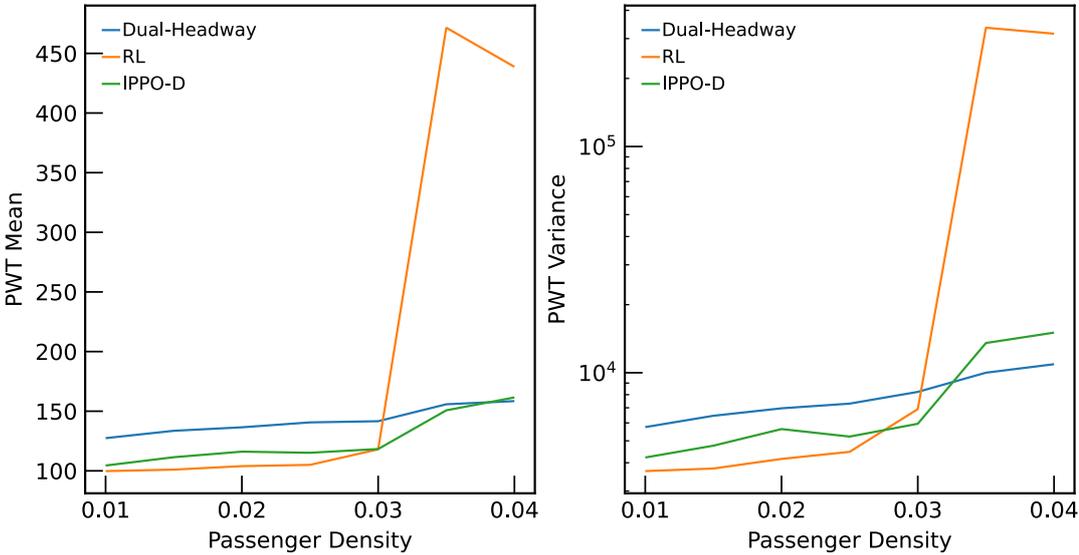}
\caption{Passenger waiting time under different control methods and environments. Passenger Density is represented by $\xi$ and passenger waiting time is in seconds. (\textbf{a}) Mean of passenger waiting time. (\textbf{b}) The standard deviation of passenger waiting time. \label{PWT}}
\end{figure}

Figure~\ref{PWT} supports our analysis of the value and limitations of different algorithms in the above two subsections. The pure RL algorithm is performing outstandingly well (with the lowest mean and variance of the passenger waiting time) in stable environments ($\xi \leq 0.03$), but in tougher environments ($\xi>0.03$), the RL network’s controlling ability is not enough to stabilize the system. Intuitively, this is because the RL network is specialized in performing accurate actions in small ranges and lacks transferability. From this result, we can infer the reason why governments are hesitating in adopting RL strategies for city bus systems. In the real world, the passenger arrival rate is constantly varying throughout the day, forming a stable environment ($\xi \leq 0.03$) most of the time but possibly creating unstable ones ($\xi > 0.03$) between morning and afternoon rush hours. It is more beneficial to keep the system stable all the time by adopting the conventional method, compared with receiving more efficiency while increasing the risk of paralyzing the system during rush hours when changing to the RL method. 

The proposed IPPO-DH method reduces the mean and variance of passenger waiting time compared with the dual-headway method and is only slightly higher compared with the RL method in stable environments. In less stable environments where the RL method fails at the control task, our method still performs with high effectiveness and preserves an efficiency comparable to conventional methods that are in use. Overall, the proposed method successfully combines the advantages of two baseline methods, creating more efficiency than the conventional method of control and a wider working range than the RL method of control.

\begin{figure}[h]
\includegraphics[width=0.9\linewidth]{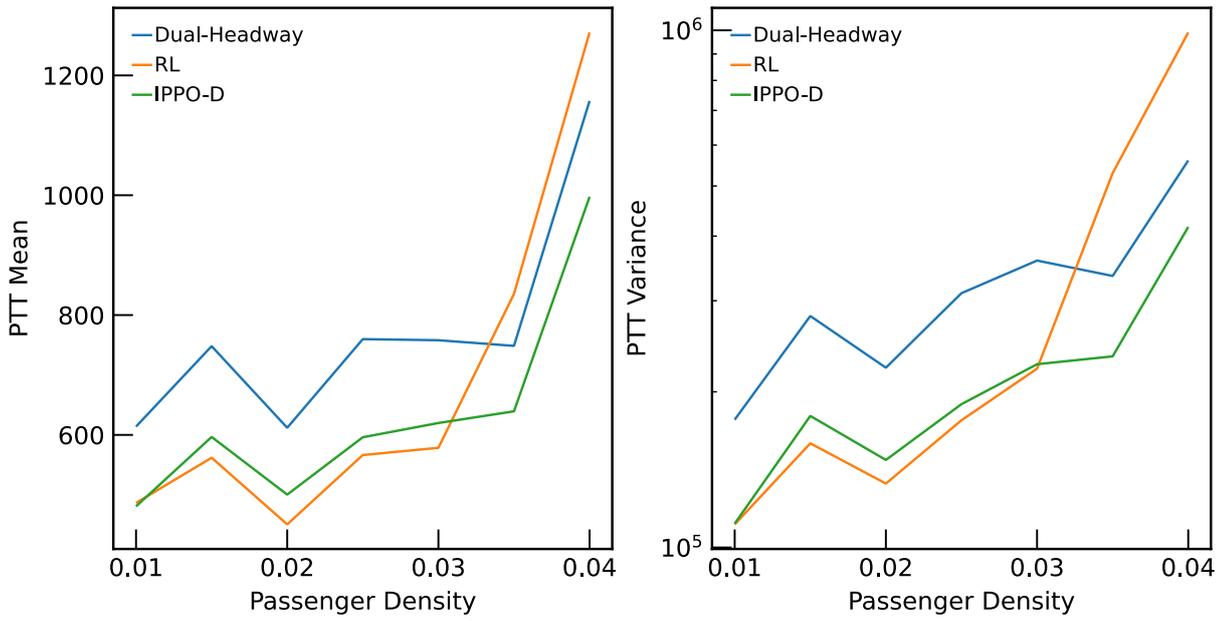}
\caption{Passenger travel time under different control methods and environments. Passenger Density is represented by $\xi$ and passenger travel time is in seconds. (\textbf{a}) Mean of passenger travel time. (\textbf{b}) The standard deviation of passenger travel time. \label{PTT}}
\end{figure}

We get a similar result on the passenger travel time (Figure~\ref{PTT}). IPPO-DH performs the task better than dual-headway control under all circumstances and greatly improves the result of RL control in harsh environments with only a slight increase in passenger travel time in stable environments.
\begin{figure}[!]
\includegraphics[width=0.9\linewidth]{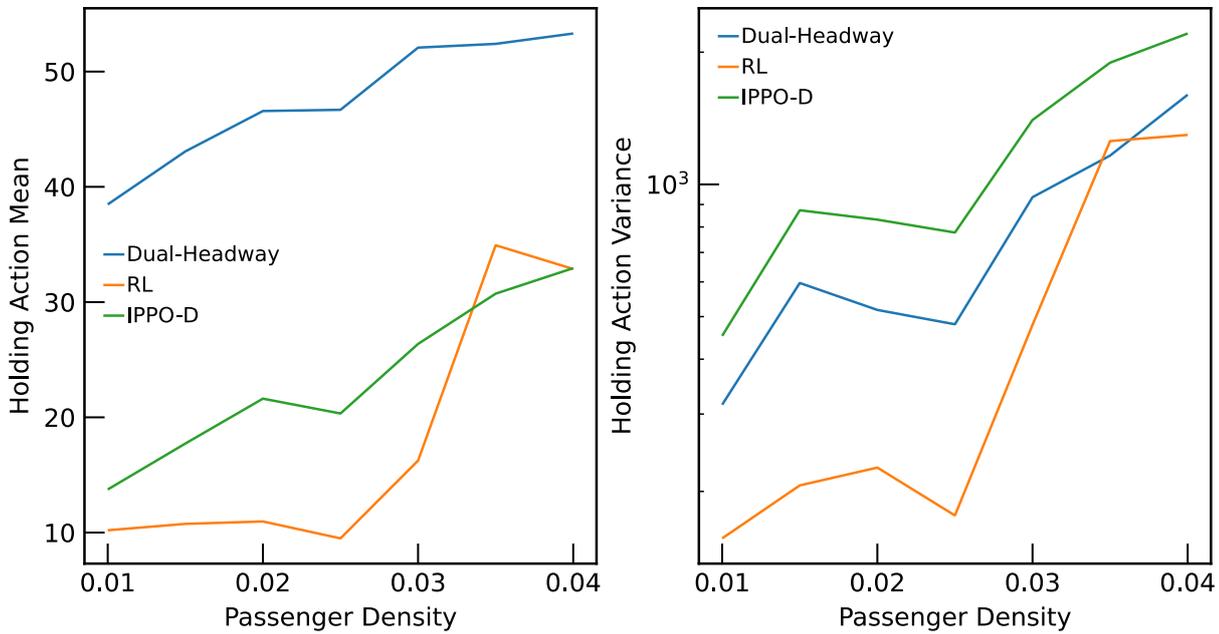}
\caption{Length of holding action under different control methods and environments. Passenger Density is represented by $\xi$ and holding action is in seconds. (\textbf{a}) Mean of holding action. (\textbf{b}) The standard deviation of holding action. \label{HC}}
\end{figure}
Interestingly, when we gather the data of holding control (Figure~\ref{HC}), we get a counter-intuitive result. Even though the mean of holding action is smaller compared to the dual-headway control method as we expected, the variance of the holding actions turns larger than both baselines in all environments. One explanation is that the delta-overlay performed by the policy network in the IPPO-DH method is reacting more sensitively to the changes in the environment to perform better in both stable and harsh environments. This causes the holding action to vary more during simulation.

%%%%%%%%%%%%%%%%%%%%%%%%%%%%%%%%%%%%%%%%%%
\section{Discussion}\label{s_D}
In this paper, to tackle the bus bunching problem, we developed a method of control to improve the efficiency and stability of existing dynamic control algorithms. More specifically, we propose an integrated control method by modifying the Dual-headway conventional control method with an RL network overlay. By experimenting with the new method in multiple simulation environments, we are confident that our methodology increases the efficiency of the bus system in common environments while maintaining effectiveness in harsh circumstances. The reduced passenger waiting and travel time along with the robustness under all circumstances brings huge application value to this new solution. 
For future work, one part is to keep updating the new method to achieve an even higher result in a common environment. Another part is to seek real-world applications to test the algorithm’s performance in real-life settings.

%%%%%%%%%%%%%%%%%%%%%%%%%%%%%%%%%%%%%%%%%%


\vspace{6pt} 

\begin{thebibliography}{999}

% Reference 1
\bibitem[Abkowitz et al.(1990)]{abkowitz1990}
Abkowitz, M.D.; Lepofsky, M. 
Implementing headway-based reliability control on transit routes.
{\em Journal of Transportation Engineering} {\bf 1990}, {\em 116}, 49––63.
% Reference 2
\bibitem[Adamski et al.(1998)]{adamski1998}
Adamski, A.; Turnau, A.
Simulation support tool for real-time dispatching control in public transport.
{\em Transportation Research Part A: Policy and
Practice} {\bf 1998}, {\em 32 (2)}, 73––87.
% Reference 3
\bibitem[Alesiani et al.(2018)]{alesiani2018}
Alesiani, F.; Gkiotsalitis, K.
Reinforcement learning-based bus holding for high-frequency services.
{\em 2018 21st International Conference on Intelligent Transportation Systems (ITSC), IEEE} {\bf 2018}, 3162-–3168.

% Reference 4
\bibitem[Barnett, A.(1974)]{barnett1974}
Barnett, A.
On controlling randomness in transit operations.
{\em Transportation Science} {\bf 1974}, {\em 8 (2)}, 102––116.
% Reference 5
\bibitem[Bartholdi et al.(2011)]{bartholdi2011}
Bartholdi, J.J.; Eisenstein, D.D.
A self-coordinating bus route to resist bus bunching.
{\em Transportation Research Part B: Methodological} {\bf 2011}, {\em 46}, 481--491.
% Reference 6
\bibitem[Berrebi et al.(2018)]{berrebi2018}
Berrebi, S.J.; Etienne Hans, E.;Chiabaut, N.; Laval, J.A.; Leclercq, L.; Watkins, K.E.
Comparing bus holding methods with and without real-time predictions.
% Reference 7
\bibitem[Boyle et al.(2009)]{boyle2009}
Boyle, D.; Pappas, J.; Boyle, P.; Nelson, B.; Sharfarz, D.; Benn, H. 
\textit{REPORT 135 Controlling System Costs: Basic and Advanced Scheduling Manuals and Contemporary Issues in Transit Scheduling.}; 
The National Academies Press: Washington, DC., 2009;

% Reference 8
\bibitem[Chen et al.(2016)]{chen2016}
Chen, W.; Zhou, K.; Chen, C. Real-time bus holding control on a transit corridor based on multi-agent reinforcement learning.
{\em 2016 IEEE 19th International Conference on Intelligent Transportation Systems (ITSC), IEEE.} {\bf 2016}, 100–-106.


% Reference 9
\bibitem[Daganzo(2009)]{dgz2009}
Daganzo, C.F. A headway-based approach to eliminate bus bunching: Systematic analysis and comparisons.
{\em Transportation Research Part B: Methodological} {\bf 2009}, {\em 43}, 913-–921.
% Reference 10
\bibitem[Daganzo et al.(2011)]{dgz2011}
Daganzo, C.F.; Pilachowski, J.
Reducing bunching with bus-to-bus cooperation. 
{\em Transportation Research Part B: Methodological} {\bf 2011}, {\em 45}, 267–-277.
% Reference 11
\bibitem[Daganzo, C.F.(1997)]{dgz1997}
Daganzo, C.F.
Schedule instability and control. In {\em Fundamentals of Transportation and Transportation Operations}; Elsevier: New York, NY, 1997; pp. 304–-309.

% Reference 12
\bibitem[Eberlein et al.(2001)]{eberlein2001}
Eberlein, X.J.; Wilson, N.H.M.; Bernstein, D.
The holding problem with real-time information available.
{\em Transportation Science} {\bf 2001}, {\em 35 (1)}, 1–-18.

% Reference 13
\bibitem[Foerster et al.(2018)]{foerster2018}
Foerster, J.N.; Farquhar, G.; Afouras, T.; Nardelli, N.; Whiteson, S.
Counterfactual multi-agent policy gradients.
{\em AAAI} {\bf 2018}, {\em 45}, 2974–-2982.

% Reference 14
\bibitem[Newell et al.(1964)]{newell964}
Newell, G.F.; Potts, R.B.
Maintaining a bus schedule. In Australian Road Research Board (ARRB) Conference, 2nd, Melbourne, Australia, 1964.

% Reference 15
\bibitem[Osuna et al.(1972)]{osuna1972}
Osuna, E.E.; Newell, G.F.
Control strategies for an idealized bus system.
{\em Transportation Science} {\bf 1972}, {\em 36 (1)}, 52––71.

% Reference 16
\bibitem[Petit et al.(2018)]{petit2018}
Petit, A.; Ouyang, Y.; Lei, C.
Dynamic bus substitution strategy for bunching intervention.
{\em Transportation Research Part B: Methodological} {\bf 2018}, {\em 115}, 1-–16.

% Reference 17
\bibitem[Schulman et al.(2017)]{schulman2017}
Schulman, J.; Wolski, F.; Dhariwal, P.; Radford, A.; Klimov, O.
Proximal policy optimization algorithms. 
{\em arXiv preprint} {\bf 2018} arXiv:1707.06347.

% Reference 18
\bibitem[Schulman et al.(2015)]{gae}
Schulman, J.; Moritz, P.; Levine, S.; Jordan, M.I.; Abbeel, P.
High-dimensional continuous control using generalized advantage estimation.
{\em arXiv preprint} {\bf 2015} arXiv:1506.02438.

{\em Transportation Research Part C: Emerging Technologies} {\bf 2018}, {\em 87}, 197--211.

% Reference 19
\bibitem[Sun et al.(2005)]{sun2005}
Sun, A.; Hickman, M.
The real–time stop–skipping problem.
{\em Journal of Intelligent Transportation Systems} {\bf 2005}, {\em 9}, 91–-109.

% Reference 20
\bibitem[Sutton et al.(1999)]{sutton1999}
Sutton, R.S.; McAllester, D.A.; Singh, S.P.; Mansour, Y.
Policy gradient methods for reinforcement learning with function approximation. In {\em Advances in Neural Information Processing systems}; S. Solla, T. Leen, K. Muller; MIT Press: Cambridge, United States, 1999; pp. 1057–-1063.
% Reference 21
\bibitem[Sutton et al.(1998)]{sutton1998}
Sutton, R.S.; Barto, A.G., et al.
\textit{Introduction to reinforcement learning. volume 2.}, 3rd ed.; MIT Press: Cambridge, United States, 1998.

% Reference 22
\bibitem[Van et al.(2010)]{van2010}
Van Oort, N.; Wilson, N.H.; Van Nes, R.
Reliability improvement in short headway transit services.
{\em Transportation Research Record} {\bf 2010}, {\em 2143}, 67--76.

% Reference 23
\bibitem[Wang et al.(2020)]{wang2020}
Wang J.; Sun L.
Dynamic holding control to avoid bus bunching: A multi-agent deep reinforcement learning framework.
{\em Transportation Research Part C} {\bf 2020}, {\em 116}, Article 102661.
% Reference 24
\bibitem[Wang et al.(2021)]{wang2021}
Wang J.; Sun L.
Reducing Bus Bunching with Asynchronous Multi-Agent Reinforcement Learning.
{\em arXiv preprint} {\bf 2021}, arXiv:2105.00376


% Reference 25
\bibitem[Xuan et al.(2011)]{xuan2011}
Xuan, Y.; Argote, J.; Daganzo, C.F.
Dynamic bus holding strategies for schedule reliability: Optimal linear control and performance analysis.
{\em Transportation Research Part B: Methodological} {\bf 2011}, {\em 45}, 1831–-1845.


% Reference 26
\bibitem[Zhao et al.(2006)]{zhao2006}
Zhao, J.; Dessouky, M.; Bukkapatnam, S.
Optimal slack time for schedule-based transit opertions.
{\em Transportation Science} {\bf 2006}, {\em 40}, 529–-539.


\end{thebibliography}
\end{document}